\documentclass{article}
\usepackage{epsf}

\begin{document}
\title{Adiabatic theorem for non-hermitian time-dependent open systems }

\author{Avner Fleischer and Nimrod Moiseyev}
\maketitle

{ Department of Chemistry and Minerva Center of Nonlinear Physics
in Complex Systems Technion -- Israel Institute of Technology
Haifa 32000, Israel. \\}

\begin{abstract}
In the conventional quantum mechanics (i.e., hermitian QM) the
adiabatic theorem for systems subjected to time periodic fields
holds only for bound systems and not for open ones (where
ionization and dissociation take place) [D. W. Hone, R. Ketzmerik,
and W. Kohn, Phys. Rev. A \textbf{56}, 4045 (1997)]. Here with the
help of the (t,t') formalism combined with the complex scaling
method we derive an adiabatic theorem for open systems and provide
an analytical criteria for the validity of the adiabatic limit.
The use of the complex scaling transformation plays a key role in
our derivation. As a numerical example we apply the adiabatic
theorem we derived to a 1D model Hamiltonian of Xe atom which
interacts with strong, monochromatic sine-square laser pulses. We
show that the generation of odd-order harmonics and the absence of
hyper-Raman lines, even when the pulses are extremely short, can
be explained with the help of the adiabatic theorem we derived.

\end{abstract}
\maketitle

{03.65.-w, 42.50.Hz, 42.65.-Ky, 32.80.Rm}

\section{Motivation}

When matter is exposed to intense laser fields, high harmonics
(HHs) of the incident radiation may be produced. Usually, only odd
harmonics are obtained even when the laser pulses are short (for
theoretical and experimental work which demonstrates this see
\cite{A-Di-Piazza+V-Veliard+S-X-Hu+S-Dionissopoulou} and
\cite{N-Hay+T-Millak}, respectively). Since the duration of the
pulse in time is inversely proportional to its width in energy
space, one may find this result surprising, as one may expect to
obtain also a large distribution of frequencies in the scattered
field. Why are only odd harmonics obtained even when the laser
pulses are short?

For CW lasers (and symmetric field-free potential) using the
non-hermitian Floquet theory it was proved that only odd harmonics
are obtained when the dynamics is controlled by a single resonance
Floquet quasienergy (QE) state \cite{Ben-Tal
Nir+Beswick+NM},\cite{Ofir+Vitali}. When laser pulses are used it
was argued that this proof holds since usually the populated
resonance states are associated with very different lifetimes and
the dynamics is controlled by the resonance state which has the
longest lifetime. However, this argument may hold only when the
duration of the laser pulses is large enough to enable the decay
of the short lived resonances. Indeed numerical simulations showed
that the harmonic generation spectra (HGS) as obtained from a
single non-hermitian (complex scaled) resonance Floquet state is
in a remarkable agreement with the results obtained from
conventional (i.e., hermitian) time dependent simulation
\cite{Nir+Kosloff+Cerjain+NM}.

The question that is addressed in this work is weather an
analytical criteria for the shape and duration of the laser pulse
for which the system is controlled by a single resonance Floquet
state can be given. It is obvious that the question regarding the
possibility of population of a single resonance state is connected
with the question regarding the degree of adiabaticity of the
process. The question is therefore under which conditions can a
short laser pulse be defined as adiabatic one. The answer to this
question is important not only to harmonic generation (HG) studies
but also for other, more general studies where lasers are used to
control the dynamics, for example adiabatic STIRAP procedures
\cite{stirap}.

In order to answer this question we use the (t,t') formalism
\cite{t-tp} together with the non-hermitian quantum mechanics
(NHQM) formalism. The use of NHQM formalism to describe the
dynamics of atoms/molecules subjected to CW laser fields is
essential, since only then can the dynamics be described in terms
physical, square integrable, resonance Floquet states. Otherwise,
the description of the dynamics in terms of hermitian Floquet
states results in very little physical insight on the problem, as
well as numerical problems, not to mention that it is limited to
the description of bound systems only. In the hermitian case the
spectrum is continuous and becomes discrete only due to the use of
finite box quantization; moreover, a single Floquet state in
hermitian QM can not describe neither the resonance phenomena nor
the field ionization phenomena \cite{Avner-Vitali-NM}.

Our strategy is as follows. In section 2 we give a brief review of
the formalism used in our derivation [namely the (t,t') formalism
for hermitian and non-hermitian Hamiltonians]. In section 3 we
introduce the new derivation of the adiabatic theorem for open
quantum systems in strong laser pulses. In section 4 we apply the
adiabatic theorem as derived in section 3 to a test-case model
hamiltonian which describes a 1D Xe atom subjected to a sin-square
pulse of monochromatic laser. In section 4 we conclude.

\section{A brief review of the \textit{(t,t')} formalism}
The (t,t') formalism enables one to obtain analytical solutions
for any time-dependent Schr\"{o}dinger equation (TDSE) with
time-dependent Hamiltonians. The formalism rests on lifting of the
TDSE to an extended Hilbert space, propagation of the wavefunction
there and finally projecting back to the physical Hilbert space.

The solution of a general TDSE is given as usual by:

\begin{equation}
H(\mathbf{r},t) \Psi(\mathbf{r},t)=i\hbar \frac{\partial}{\partial
t} \Psi(\mathbf{r},t)\label{eq1}
\end{equation}

with the initial condition $\Psi(\mathbf{r},t=0)$ and the vector
operator $\mathbf{r}$ describes the internal degrees of freedom.
It has been shown by Peskin and Moiseyev that by regarding time as
an extra coordinate $t'$, one can obtain another Schr\"{o}dinger
equation with a time independent Hamiltonian in the extended
Hilbert space $(\mathbf{r},t')$, whose solution
$\overline{\Psi}(\mathbf{r},t',t)$ has an analytical time
dependence (given by the analytical time evolution operator
associated with time-independent Hamiltonians) \cite{t-tp}. Our
desired wavfunction $\Psi(\mathbf{r},t)$ can be deduced from this
wavfunction by a simple operation (that will be shown later). The
advantage is that efficient propagation schemes designed for
solving the TDSE with time-independent Hamiltonians could then be
used also for time dependent Hamiltonians, thus releasing one from
the difficulties associated with time ordering.

Let us define the following Floquet-type operator

\begin{equation}
H_{F}(\mathbf{r},t') \equiv (H(\mathbf{r},t)-i\hbar
\frac{\partial}{\partial t})|_{t'=t}=H(\mathbf{r},t')-i\hbar
\frac{\partial}{\partial t'}~,~0\leq t' \leq t_{f}
 \label{eq2}
\end{equation}
where $t'$ should be regarded as a {\textbf{coordinate}} and
$t_{f}$ is the final time of propagation. When the Hamiltonian
$H(\mathbf{r},t)$ is time periodic with period $T$, the operator
is the Floquet operator and $t_{f}\equiv T$. Provided that correct
boundary conditions are chosen for the $t'$ coordinate, this
operator is hermitian since it is the sum of two hermitian
operators. This operator has eigenstates and eigenvalues which are
given by the eigenvalue equation:

\begin{equation}
H_{F}(\mathbf{r},t')
\phi_{\alpha}(\mathbf{r},t')=\varepsilon_{\alpha}
\phi_{\alpha}(\mathbf{r},t')
 \label{eq3}
\end{equation}
and the set of eigenstates is complete in the extended Hilbert
space $(\mathbf{r},t')$ with respect to the inner product

\begin{equation}
\langle \langle \phi_{\alpha}|\phi_{\alpha'} \rangle\rangle
_{\mathbf{r},t'}\equiv
\frac{1}{t_{f}}\int_{0}^{T_{p}}dt'\int_{-\infty}^{\infty}d^{3}r
\phi^{*}_{\alpha}(\mathbf{r},t') \phi_{\alpha'}(\mathbf{r},t')
=\delta_{\alpha,\alpha'}
 \label{eq4}
\end{equation}

Say the following TDSE with time-independent Hamiltonian need to
be solved with the initial state
$\overline{\Psi}(\mathbf{r},t',t=0)$:

\begin{equation}
H_{F}(\mathbf{r},t') \overline{\Psi}(\mathbf{r},t',t)=i\hbar
\frac{\partial}{\partial t}
\overline{\Psi}(\mathbf{r},t',t)\label{eq5}
\end{equation}
Using the definition of $H_{F}(\mathbf{r},t')$ this equation reads

\begin{equation}
H(\mathbf{r},t') \overline{\Psi}(\mathbf{r},t',t)=i\hbar
(\frac{\partial}{\partial t}+\frac{\partial}{\partial t'})
\overline{\Psi}(\mathbf{r},t',t)\label{eq6}
\end{equation}
By setting the cut $t=t'$ on Eq.\ref{eq6} one gets:

\begin{equation}
[H(\mathbf{r},t')
\overline{\Psi}(\mathbf{r},t',t)]|_{t'=t}=H(\mathbf{r},t)
[\overline{\Psi}(\mathbf{r},t',t)|_{t'=t}]=i\hbar
\frac{\partial}{\partial t}
[\overline{\Psi}(\mathbf{r},t',t)|_{t'=t}]\label{eq7}
\end{equation}
where the following property

\begin{equation}
\frac{\partial}{\partial t}
[\overline{\Psi}(\mathbf{r},t',t)|_{t'=t}]=
[(\frac{\partial}{\partial t}+\frac{\partial}{\partial t'})
\overline{\Psi}(\mathbf{r},t',t)]|_{t'=t}\label{eq8}
\end{equation}
(which holds true for \textbf{any} function of $t'$ and $t$. The
relation between the solution of Eq.\ref{eq7} and the solution of
the original TDSE (Eq.\ref{eq1}) is given by

\begin{equation}
\Psi(\mathbf{r},t)=\overline{\Psi}(\mathbf{r},t',t)|_{t'=t}\label{eq9}
\end{equation}
provided that these two differential equations have the same
initial condition. Hence,

\begin{equation}
\overline{\Psi}(\mathbf{r},t',t)|_{t'=t=0}=\Psi(\mathbf{r},t=0)\label{eq10}
\end{equation}

It is seen from Eq.\ref{eq10} that apparently the initial
condition isn't unique. While this holds true in case that one is
interested only in the physical wavefunction $\Psi(\mathbf{r},t)$,
it should be noted that if one wishes to calculate physical
quantities in the extended Hilbert space using the function
$\overline{\Psi}(\mathbf{r},t',t)$ and then go back to the
original Hilbert space, and get the correct results, the initial
condition in the extended Hilbert space should behaves as a
delta-function in $t'$. The correct initial condition will
therefore be
$\overline{\Psi}(\mathbf{r},t',t=0)=\Psi(\mathbf{r},t=0)\delta(t')$.

The main advantage of the (t,t') formalism from a numerical point
of view is that it enables the use of an analytical expression for
the time evolution operator, without the necessity of time
ordering, even when the Hamiltonian is strongly time-dependent.
Any TDSE with time-dependent Hamiltonian could be replaced by a
different TDSE, with time-independent Hamiltonian, for which a
greater number of accurate integration schemes exist and the
solution is given formally by
$\overline{\Psi}(\mathbf{r},t',t)=e^{-\frac{i}{\hbar}H_{F}(\mathbf{r},t')
t}\overline{\Psi}(\mathbf{r},t',t=0)$). The price one pays however
is that the new TDSE need to be integrated over one more
dimension.

The main advantage of the \textit{(t,t')} formalism from a
conceptional point of view is that it enables to describe any
time-dependent dynamics in terms of stationary eigenstates and
eigenvalues.

So far the derivation has been carried out within the framework of
the conventional (i.e., hermitian) quantum mechanics. Using the
complex-scaling (CS) transformation \cite{Reinhardt,Reviewnimrod}

\begin{equation}
H_{F}(\mathbf{r},t') \longrightarrow
H^{\theta}_{F}(\mathbf{r},t')\equiv
H(\mathbf{r}e^{i\theta},t)-i\hbar \frac{\partial}{\partial t}
\label{eq10b}
\end{equation}
the quasienergy spectrum of the Floquet Hamiltonian becomes
complex and square-integrable resonance states, that were embedded
in the continuum in the unscaled problem, are uncovered. For sake
of simplicity we drop the index $\theta$ in all $\theta$-dependent
expressions (operators, eigenvalues, eigenvectors etc.) in the
proceeding text.

The Floquet operator $H_F$ can be represented with the orthogonal
Fourier basis set $\frac{1}{\sqrt{T}} e^{i\omega n t'}, n=0, \pm1,
\pm2,...$ as a square matrix

\begin{equation}
[\underline{\underline{H_F(\mathbf{r})}}]_{n',n}=\frac{1}{T}\int_{0}^{T}dt'e^{-i\omega
n' t'} H_F(\mathbf{r},t')e^{-i\omega n t} \label{eq11}
\end{equation}
where $T=2\pi/\omega$. The left and right eigenvectors
(bi-orthogonal set of eigenvectors, \cite{Wilkinson}) of this
Floquet matrix are the Fourier components of the Floquet states as
defined in Eq.\ref{eq3}, that is

\begin{equation}
[\underline{\underline{H_F(\mathbf{r})}}]^{t}\overrightarrow{\varphi}^{L}_{\alpha}(\textbf{r})=E_{\alpha}\overrightarrow{\varphi}^{L}_{\alpha}(\textbf{r})
\label{eq11a}
\end{equation}

\begin{equation}
[\underline{\underline{H_F(\mathbf{r})}}]\overrightarrow{\varphi}^{R}_{\alpha}(\textbf{r})=E_{\alpha}\overrightarrow{\varphi}^{R}_{\alpha}(\textbf{r})
\label{eq11b}
\end{equation}
where

\begin{equation}
[\overrightarrow{\varphi}^{L/R}_{\alpha}(\textbf{r})]_{n}\equiv\varphi^{L/R}_{\alpha,n}(\textbf{r}).
\label{eq11c}
\end{equation}

Since in our case
$[\underline{\underline{H_F(\mathbf{r})}}]^{t}=[\underline{\underline{H_F(\mathbf{r})}}]$
then
$\overrightarrow{\varphi}^{L}_{\alpha}(\textbf{r})=\overrightarrow{\varphi}^{R}_{\alpha}(\textbf{r})$.
There are two sets of eigenfunctions of the Floquet operator
$H_F(\mathbf{r},t')$

\begin{equation}
\phi^{R}_\alpha(\textbf{r},t')=\sum_n
\varphi^{R}_{\alpha,n}(\textbf{r})e^{i\omega nt'} \label{eq12}
\end{equation}
and

\begin{equation}
\phi^{L}_\alpha(\textbf{r},t')=\sum_n
\varphi^{L}_{\alpha,n}(\textbf{r})e^{-i\omega nt'}=\sum_n
\varphi^{R}_{\alpha,n}(\textbf{r})e^{-i\omega nt'} \label{eq13}
\end{equation}

As pointed out in \cite{NM-Certain-Weinhold-1978} the c-product,
which is associated with the non usual inner-product in linear
algebra (see for example Wilkinson's text book \cite{Wilkinson}),
reads

\begin{equation}
(\phi^{L}_{\alpha}(\textbf{r},t')|\phi^{R}_{\alpha'}(\textbf{r},t'))_{\mathbf{r},t'}=\sum_n
(\varphi^{R}_{\alpha,n}|\varphi^{R}_{\alpha',n})_{\mathbf{r}}=\delta_{\alpha,\alpha'}
\label{eq14}
\end{equation}
where
$(\phi^{L}_{\alpha}(\textbf{r},t')|\phi^{R}_{\alpha'}(\textbf{r},t'))_{\mathbf{r},t'}\equiv
\langle
\phi^{L*}_{\alpha}(\textbf{r},t')|\phi^{R}_{\alpha'}(\textbf{r},t')
\rangle_{\mathbf{r},t'}$. However, as proposed recently by
Moiseyev and Lein (\cite{F-product}), the inner-product should be
modified even further when time-dependent functions are used as
basis set due to the time-asymmetry problem in NHQM. If we define
$\phi^{R}_\alpha(\textbf{r},t)$ and
$E_{\alpha}=E_{r_{\alpha}}-\frac{i}{2}\Gamma_{\alpha}$ to be the
Floquet eigenfunctions and eigenvalues respectively, then the
time-dependent basis functions could be

\begin{equation}
\Psi^{R}_{\alpha}(\textbf{r},t',t)=e^{-\frac{i}{\hbar}E_{\alpha}t}\phi^{R}_{\alpha}(\textbf{r},t')
\label{eq14a}
\end{equation}
(which are solutions of the TDSE) and also the left functions

\begin{equation}
\Psi^{L}_{\alpha}(\textbf{r},t',t)=e^{+\frac{i}{\hbar}E^{*}_{\alpha}t}\phi^{L}_{\alpha}(\textbf{r},t').
\label{eq14b}
\end{equation}
Following the modified definition of the inner product
("finite-range" product, "F-product" \cite{F-product})

\begin{equation}
(\Psi^{L}_{\alpha}(t)|\Psi^{R}_{\alpha'}(t))_{\mathbf{r},t'}\equiv
e^{-\frac{i}{\hbar}E_{\alpha}t}e^{+\frac{i}{\hbar}
E^{*}_{\alpha'}t}(\phi^{L}_{\alpha}(\textbf{r},t')|\phi^{R}_{\alpha'}(\textbf{r},t'))_{\mathbf{r},t'}
=e^{-\Gamma_\alpha t}\delta_{\alpha,\alpha '} \label{eq15}
\end{equation}
it implies that the $\alpha^{th}$ quasienergy state decays
exponentially in time. For more detailed discussion see
\cite{F-product},\cite{F2-product}.

\section{The adiabatic theorem for time-dependent systems} The
adiabatic theorem for time-dependent \textbf{bound} systems was
derived in 1997 by Kohn \textit{et al.} \cite{Kohn} and in 1999 by
Holthaus \textit{et al.} \cite{K.Drese-M.Holthaus} who used the
(t,t') formalism to describe the evolution of a system subjected
to chirped laser pulses. As discussed by Kohn and co-workers, the
adiabatic approach is \textbf{not} applicable for open systems
since the quasienergy level spacing reduce to zero as the number
of basis functions used in the numerical calculation is increased.
To avoid this difficulty Baer et al. \cite{Baer} applied the
adiabatic theorem to time-dependent \textbf{open} systems in the
high frequency regime where the system was stabilized and the
resonances (which were embedded in the continuum part of the
Floquet spectra) became so narrow that they could be practically
treated as bound states. The purpose of our work is to derive the
adiabatic theorem for general time-dependent open systems where
there are no bound states and the resonances aren't necessarily
narrow. We are using the non-hermitian Floquet formalism (through
the CS formalism) which allow us to describe the dynamics in term
of non-hermitian resonance states (see for example
\cite{Reinhardt}-\cite{Potvliege}, and also the work of Day et al.
\cite{Day} who used the NH Floquet multistate method to study the
applicability of the single Floquet resonance approximation in the
description of the dynamics of H atom subjected to intense laser
fields of various strengths).

Below we derive the adiabatic theorem for time-dependent open
systems using the extended (t,t') formalism. By the term
"extended" \textit{(t,t')} formalism we mean that in the same
manner presented, one may add any number of time "coordinates" to
the Schr\"{o}dinger equation as one wishes, if by this a better
understanding or easier solution of the problem is achieved. Here
we found that by addition of 2 time "coordiantes" to the TDSE, we
simplified the derivation of the adiabaticity criteria for
photo-induced dynamical systems. In this sense, we are using a
\textit{(t,t',t'')} formalism.

We would like to study the dynamics of a single active electron in
an atom or molecule, subjected to a pulse of strong monochromatic
linearly polarized laser radiation. In the dipole approximation
the TDSE which describes this process is:

\begin{equation}
H(\mathbf{r},t) \Psi^{R}(\mathbf{r},t)=i\hbar
\frac{\partial}{\partial t} \Psi^{R}(\mathbf{r},t)\label{eq16}
\end{equation}
where,

\begin{equation}
H(\mathbf{r},t)=H_{0}(\mathbf{r})-e \mathbf{r} \cdot \mathbf{f}(t)
cos(\omega t) \label{eq17}
\end{equation}
and,

\begin{equation}
\mathbf{f}(t)\equiv  \varepsilon_{0} \mathbf{e_{k}} f(t).
\label{eq18}
\end{equation}
Here $f(t)$ is the function which describes the envelope of the
laser pulse and $\mathbf{f}(t)$ is a vector as defined in
Eq.\ref{eq18}; $\varepsilon_{0}$ is the laser's amplitude,
$\mathbf{e_{k}}$ is a unit vector in the direction of the electric
component of the laser field, $\omega$ is the laser's frequency,
with $T=2 \pi/ \omega$ the optical period. $H_{0}(\mathbf{r})$ is
the field-free Hamiltonian and the vector operator $\mathbf{r}$
describes the internal degrees of freedom (the coordinates are
complex-scaled throughout) .

In the same spirit of subsection (2.1), we define the following
operator

\begin{equation}
H_{F}(\mathbf{r},t',t'') \equiv
\widetilde{H}(\mathbf{r},t',t'')-i\hbar \frac{\partial}{\partial
t'}-i\hbar \frac{\partial}{\partial t''}
 \label{eq19}
\end{equation}
where

\begin{equation}
\widetilde{H}(\mathbf{r},t',t'') \equiv H_{0}(\mathbf{r})-e
\mathbf{r} \cdot \mathbf{f}(t'') cos(\omega t')
 \label{eq20}
\end{equation}
and $t',t''$ should be regarded as additional
\textbf{coordinates}. Upon complex-scaling \cite{cs}
$H_{F}(\mathbf{r},t',t'')$ becomes non-hermitian. Therefore the
inner c-product should be used as mentioned before. The
quasi-energy solutions of this complex-scaled Floquet-type
operator are:

\begin{equation}
H_{F}(\mathbf{r},t',t'') \psi^{R}_{k}(\mathbf{r},t',t'')=\lambda
_{k} \psi^{R}_{k}(\mathbf{r},t',t'')
 \label{eq21}
\end{equation}

\begin{equation}
H^{\dag*}_{F}(\mathbf{r},t',t'')
\psi^{L}_{k}(\mathbf{r},t',t'')=\lambda _{k}
\psi^{L}_{k}(\mathbf{r},t',t'')
 \label{eq22}
\end{equation}
where the symbol $H^{\dag*}_{F}$ doesn't stand for an operator but
for the transpose of the matrix representing the operator $H_{F}$.
The eigenfunctions form a complete set in the extended Hilbert
space $\mathbf{r},t',t''$.

Say we want to solve the following TDSE with the initial state
$\widetilde{\Psi}(\mathbf{r},t',t'',t=0)$:

\begin{equation}
H_{F}(\mathbf{r},t',t'')
\widetilde{\Psi}^{R}(\mathbf{r},t',t'',t)=i\hbar
\frac{\partial}{\partial t}
\widetilde{\Psi}^{R}(\mathbf{r},t',t'',t)\label{eq23}
\end{equation}
The solution to this equation is

\begin{equation}
\widetilde{\Psi}^{R}(\mathbf{r},t',t'',t)=e^{-\frac{i}{\hbar}H_{F}(\mathbf{r},t',t'')t}\widetilde{\Psi}^{R}(\mathbf{r},t',t'',t=0)
=\sum_{k}c_{k} e^{-\frac{i}{\hbar}\lambda _{k}t}
\psi^{R}_{k}(\mathbf{r},t',t'') \label{eq24}
\end{equation}
and a function $\widetilde{\Psi}^{L}(\mathbf{r},t',t'',t)$, which
is \textbf{not} a solution of any Schr\"{o}dinger equation, is
defined as

\begin{equation}
\widetilde{\Psi}^{L}(\mathbf{r},t',t'',t)=\sum_{k}c_{k}
e^{+\frac{i}{\hbar}\lambda^{*}_{k}t}
\psi^{L}_{k}(\mathbf{r},t',t'') \label{eq25}
\end{equation}
where by taking the cut $t'=t''=t=0$ on Eq.\ref{eq24} it is easily
seen that the expansion coefficients are
$c_{k}=(\psi^{L}_{k}(\mathbf{r},t',t'')|_{t'=t''=0}|\widetilde{\Psi}^{R}(\mathbf{r},t',t'',t)|_{t'=t''=t=0})_{\mathbf{r}}$.
[$e^{-\frac{i}{\hbar}\lambda _{k}t}
\psi^{R}_{k}(\mathbf{r},t',t'')|_{t'=t''=t}$ is a solution of the
original TDSE (apply the cut t'=t''=t on Eq.\ref{eq21} and compare
the result to the result obtained when the function
$e^{-\frac{i}{\hbar}\lambda _{k}t}
\psi^{R}_{k}(\mathbf{r},t',t'')|_{t'=t''=t}$ is substituted in the
original TDSE); therefore, any linear combination of these
solutions is also a solution].

Let us now return to the main purpose of this article, the
derivation of the adiabatic theorem for non-hermitian open
systems. We would like to treat $t''$ as an adiabatic coordinate
(this is the "coordinate" associated with the pulse envelope) in
the same way that the electronic motion is separated from the
nuclear one in the treatment of molecules within the
Born-Oppenheimer approximation. First we define the following
operator

\begin{equation}
H_{ad}(\mathbf{r},t',t'') \equiv
\widetilde{H}(\mathbf{r},t',t'')-i\hbar \frac{\partial}{\partial
t'}
 \label{eq26}
\end{equation}
where $t''$ should be regarded as a parameter now. This means that
this Hamiltonian is a Floquet Hamiltonian describing the
interaction of the atom with \textbf{CW} laser of strength
$\varepsilon_{1}$ where following Eq.\ref{eq18}

\begin{equation}
\varepsilon_{1}=|\mathbf{f}(t)|=\varepsilon_{0}f(t)
 \label{eq26a}
\end{equation}

The eigenstates of this operator form a complete basis (in the
$\mathbf{r}-t'$ space), for every value of the parameter $t''$:

\begin{equation}
H_{ad}(\mathbf{r},t',t'')
\psi^{ad,R}_{\alpha}(\mathbf{r},t',t'')=\varepsilon^{ad}_{\alpha}(t'')
\psi^{ad,R}_{\alpha}(\mathbf{r},t',t'').
 \label{eq27}
\end{equation}
Notice that due to the complex scaling, $\lambda _{k}$ and
$\varepsilon^{ad}_{\alpha}(t'')$ get complex values.

We can expand each eigenstate of the complete problem
(Eq.\ref{eq21}) in this basis:

\begin{equation}
\psi^{R}_{k}(\mathbf{r},t',t'')=\sum_{\alpha'}\psi^{ad,R}_{\alpha'}(\mathbf{r},t',t'')
\chi_{\alpha',k}(t'')\label{eq28}
\end{equation}
Substituting Eq.\ref{eq28} into Eq.\ref{eq21}, multiplying the
obtained equation from the left hand side by
$\psi^{ad,L}_{\alpha'}(\mathbf{r},t',t'')$ and integrating over
$\mathbf{r}$ and $t'$ one gets, in matrix notation, the equality:

\begin{equation}
[-i\hbar \frac{\partial}{\partial t''}
\underline{\underline{I}}+(\underline{\underline{E}}^{ad}(t'')+\underline{\underline{V}}(t''))]\overrightarrow{\chi}_{k}(t'')=\lambda_{k}\overrightarrow{\chi}_{k}(t'')
\label{eq29}
\end{equation}
where

$$
[\underline{\underline{E}}^{ad}(t'')]_{\alpha,\alpha'}=\varepsilon^{ad}_{\alpha}(t'')\delta_{\alpha,\alpha'},
$$
$$
[\underline{\underline{V}}(t''))]_{\alpha,\alpha'}=((\psi^{ad,L}_{\alpha}(\mathbf{r},t',t'')|-i\hbar
\frac{\partial}{\partial t''}
|\psi^{ad,R}_{\alpha'}(\mathbf{r},t',t'')))_{\mathbf{r},t'}
$$

\begin{equation}
[\overrightarrow{\chi}_{k}(t'')]_{\alpha}=\chi_{\alpha,k}(t'').
\label{eq30}
\end{equation}
Notice that in case that the matrix on the left hand side of
Eq.\ref{eq29} is diagonal, a homogeneous systems of uncoupled
equations is obtained. In such a case one should solve each
equation separately. Therefore, the sum in Eq.\ref{eq28} reduces
to a single product. This is exactly the adiabatic approximation
as appears in the Born-Oppenheimer context. The next step in our
derivation is to represent the matrix
$(\underline{\underline{E}}^{ad}(t'')+\underline{\underline{V}}(t''))$
by its spectral decomposition

\begin{equation}
[\underline{\underline{E}}^{ad}(t'')+\underline{\underline{V}}(t'')]\underline{\underline{D}}^{R}(t'')=\underline{\underline{D}}^{R}(t'')\underline{\underline{W}}(t'')
\label{eq31}
\end{equation}

\begin{equation}
[\underline{\underline{E}}^{ad}(t'')+\underline{\underline{V}}(t'')]^{t}\underline{\underline{D}}^{L}(t'')=\underline{\underline{D}}^{L}(t'')\underline{\underline{W}}(t'')
\label{eq32}
\end{equation}
The matrix of eigenvalues $\underline{\underline{W}}(t'')$ is
diagonal and the right and left eigenvectors are normalized with
respect to each other in order to maintain the correct inner
product:

\begin{equation}
[\underline{\underline{D}}^{L}(t'')]^{t}\underline{\underline{D}}^{R}(t'')=\underline{\underline{I}}
\label{eq33}
\end{equation}

In the case that the matrix
$(\underline{\underline{E}}^{ad}(t'')+\underline{\underline{V}}(t''))$
is not strictly diagonal we can use first-order perturbation
theory to get the first-order deviation from diagonal. If we treat
the matrix $\underline{\underline{V}}(t'')$ as perturbation, we
get

\begin{equation}
[\underline{\underline{D}}^{R}(t'')]_{\alpha',\alpha}=\delta_{\alpha',\alpha}+\frac{[\underline{\underline{V}}(t'')]_{\alpha',\alpha}}{\varepsilon^{ad}_{\alpha}(t'')-\varepsilon^{ad}_{\alpha'}(t'')}
\label{eq34}
\end{equation}
The matrix will be diagonal to a good approximation if

\begin{equation}
A_{\alpha}(t'')\equiv
\sum_{\alpha'\neq\alpha}|[\underline{\underline{D}}^{R}(t'')]_{\alpha',\alpha}|\ll
1 \label{eq35}
\end{equation}
which produces the following adiabaticity criteria,

\begin{equation}
A_{\alpha}(t'')\equiv
\sum_{\alpha'\neq\alpha}|\frac{((\psi^{ad}_{\alpha}(\mathbf{r},t',t'')|-i\hbar
\frac{\partial}{\partial t''}
|\psi^{ad}_{\alpha'}(\mathbf{r},t',t'')))_{\mathbf{r},t'}}{\varepsilon^{ad}_{\alpha'}(t'')-\varepsilon^{ad}_{\alpha}(t'')}|\ll
1 \label{eq36}
\end{equation}

Using the specific form of the Hamiltonian of the problem given in
Eq.\ref{eq17} and the Hellman-Feynman theorem one gets the
adiabatic condition for time-dependent open systems:

\begin{equation}
A_{\alpha}(t'')\ll 1 \label{eq36a}
\end{equation}
where

\begin{equation}
A_{\alpha}(t'')\equiv |e|\hbar \varepsilon_{0} |\frac{d f(t'') }{d
t''}|
\sum_{\alpha'\neq\alpha}|\frac{((\psi^{ad,L}_{\alpha}(\mathbf{r},t',t'')|
\mathbf{r}\cdot \mathbf{e_{k}} \cos(\omega t')
|\psi^{ad,R}_{\alpha'}(\mathbf{r},t',t'')))_{\mathbf{r},t'}}{(\varepsilon^{ad}_{\alpha'}(t'')-\varepsilon^{ad}_{\alpha}(t''))^{2}}|
\label{eq37}
\end{equation}

The index $\alpha$ is a super-index; since (Eq.\ref{eq27}) it is
easily seen that not only is
$\psi^{ad,L/R}_{\alpha}(\mathbf{r},t',t'')$ a solution of the
eigenvalue equation, with eigenvalue
$\varepsilon^{ad}_{\alpha}(t'')$, but also $e^{i\omega mt'
}\psi^{ad,L/R}_{\alpha}(\mathbf{r},t',t'')$ is a solution, with
the eigenvalue $\varepsilon^{ad}_{\alpha}(t'')+\hbar\omega m$, for
any integer $m$. Let us take all the states whose corresponding
eigenvalues lie in the interval $[0,\hbar\omega]$ (the first
Brillouin zone) and define them to have index $m=0$; we call these
states $\phi^{ad,L}_{j}(\mathbf{r},t',t'')$,
$\phi^{ad,R}_{j}(\mathbf{r},t',t'')$ and the corresponding
eigenvalues $E^{ad}_{j}(t'')$ and get:

\begin{equation}
\psi^{ad,R}_{\alpha}(\mathbf{r},t',t'')\equiv
\phi^{ad,R}_{j}(\mathbf{r},t',t'')e^{i\omega mt'} \label{eq38}
\end{equation}

\begin{equation}
\psi^{ad,L}_{\alpha}(\mathbf{r},t',t'')\equiv
\phi^{ad,L}_{j}(\mathbf{r},t',t'')e^{i\omega mt'} \label{eq39}
\end{equation}

\begin{equation}
\varepsilon^{ad}_{\alpha}(t'')\equiv E^{ad}_{j}(t'')+\hbar\omega
m\label{eq40}
\end{equation}
where

\begin{equation}
0\leq E^{ad}_{j}(t'')\leq \hbar\omega \label{eq41}
\end{equation}

Thus, the index $\alpha$ actually counts both the position of the
quasienergy within the first Brillouin zone (the index $j$) and
the Brillouin zone itself (the index $n$). With respect to the
generalized inner product, two states with one or more of the
indices ($j$,$n$) different are orthogonal.

Going back to Eq.\ref{eq37} now ,it is seen that the probability
to couple an initial adiabatic state
$\psi^{ad}_{\alpha=(j,0)}(\mathbf{r},t',t'')$ to any other
adiabatic state $\psi^{ad}_{\alpha'=(j',m')}(\mathbf{r},t',t'')$
is given by,

\begin{equation}
A_{(j,0)}(t'')=F(t'')\cdot \sum_{j'\neq j}a_{(j')}^{(j)}(t'')
\label{eq42}
\end{equation}
where

\begin{equation}
F(t'')=|e|\hbar \varepsilon_{0}|\frac{d f(t'') }{d t''}|,
\label{eq43}
\end{equation}

\begin{equation}
a_{(j')}^{(j)}(t'')=\sum_{m\neq 0}|c_{(j',m)}^{(j,0)}(t'')|
\label{eq44}
\end{equation}
and the functions $c_{(j',m)}^{(j,0)}(t'')$, that will be termed
as \textit{"adiabatic cross terms"} from now on, are given by

\begin{equation}
c_{(j',m)}^{(j,0)}(t'')=
\frac{((\phi^{ad,L}_{j}(\mathbf{r},t',t'')|\textbf{r}\cdot
\mathbf{e_{k}} cos(\omega t') e^{i\omega mt'}
|\phi^{ad,R}_{j'}(\mathbf{r},t',t'')))_{\mathbf{r},t'}}{(E^{ad}_{j'}(t'')-E^{ad}_{j}(t'')+\hbar\omega
m )^{2}}\label{eq45}
\end{equation}

Since the energies $E^{ad}_{j'}(t'')$ are complex (the Hamiltonian
is non-hermitian) then for $j\neq j'$ it is most unlikely that
$E^{ad}_{j'}=E^{ad}_{j}$. It is clear that the denominator hardly
ever vanishes even when $m=0$. This holds true even when $j=j'$
but $m\neq 0$.

\textit{Therefore, the criteria for a pulse to be considered
adiabatic is that the condition}

\begin{equation}
A_{(j,0)}(t'')\ll 1 \label{eq46}
\end{equation}

\textit{be fulfilled.}

Who is the adiabatic state
$\psi^{ad}_{\alpha=(j,0)}(\mathbf{r},t',t'')$ whose couplings to
all other states $\psi^{ad}_{\alpha'=(j',m')}(\mathbf{r},t',t'')$
should remain small in the adiabatic limit? Assuming that the
system is in a stationary state of the field-free problem
$\varphi_{j}(\textbf{r})$ before the action of the field (relaying
on the superposition principle of the solutions of the TDSE
generality is not lost by this assumption) and provided that the
field is switched adiabatically, this Floquet resonanace state
$\psi^{ad}_{\alpha=(j,0)}(\mathbf{r},t',t'')$ is the state which
is "born" from the stationary state $\varphi_{j}(\textbf{r})$ as
the field is switched on. If the process is not done adiabatically
many Floquet resonanace states will be populated, resulting in
considerable couplings of
$\psi^{ad}_{\alpha=(j,0)}(\mathbf{r},t',t'')$ to them and collapse
of the adiabatic condition in Eq.\ref{eq46}. The only adiabatic
check which is physically meaningful is one in which
\textit{$\alpha$ denotes a resonance state} (which is associated
with square-integrable function). $\alpha'$ however \textit{stands
for both resonances and rotated continuum states}.

One should notice that the derivation of the adiabatic theorem
presented above holds for many electron systems. In order to avoid
complicated notation the symbol $\mathbf{r}$ can stand for many
electrons. It also holds for polychromatic radiation, where the CW
field is a collection of monochromatic fields with frequencies
$\omega_{i}$ and phases $\varphi_{i}$.

Notice that for a given problem (given spectral profile of the CW
field and given field-free potential), the sum over absolute value
of the adiabatic cross terms should be calculated as function of
the effective CW-field intensity (which is symbolized here through
$t''$) \textbf{only once}. The adiabatic cross terms should then
be converted to be functions of time, through the explicit time
dependence of the pulse envelope and then the sum of their
absolute values should be multiplied by the time derivative of the
pulse envelope and by the maximal field intensity to obtain the
final expression which indicates whether the adiabatic criteria is
fulfilled or not.

It is easily seen in Eq.\ref{eq43} that for a given system, the
shape and intensity of the laser pulse determines its adiabaticity
since these parameters influence the shape-derivative term. A
short pulse which is switched on or off abruptly and has a high
maximal intensity, will most likely not be adiabatic.

In the case that the adiabaticity criteria is fulfilled, the sum
in Eq.\ref{eq28} could be reduced to a single product:

\begin{equation}
\psi_{k,j}^{R}(\mathbf{r},t',t'')\cong
\phi^{ad,R}_{j}(\mathbf{r},t',t'') \chi_{j,k}(t'')\label{eq51}
\end{equation}
and the adiabatic states are assigned with two good quantum
numbers $k$ and $j$. The solution to the eigenvalue equation
Eq.\ref{eq29} is now
($[\underline{\underline{V}}(t'')]_{\alpha,\alpha'}\approx 0$)

\begin{equation}
\chi_{j,k}(t'')=e^{+\frac{i}{\hbar}\lambda_{k}t''}e^{-\frac{i}{\hbar}\int^{t''}d\tau
E^{ad}_{j}(\tau)}. \label{eq52}
\end{equation}
By using Eq.\ref{eq24},\ref{eq51},\ref{eq52} we get that the
adiabatic solution, in the (t,t',t'') formalism, is given by,

\begin{equation}
\widetilde{\Psi}^{R}(\mathbf{r},t',t'',t)=[\sum_{k}c_{k}e^{-\frac{i}{\hbar}\lambda
_{k}(t-t'')}]\phi^{ad,R}_{j}(\mathbf{r},t',t'')e^{-\frac{i}{\hbar}\int^{t''}d\tau
E^{ad}_{j}(\tau)}\label{eq53}
\end{equation}
Now, applying the cut $t''=t$ in order to obtain the physical
solution of Eq.\ref{eq16} one gets (eliminating the phase factor
$\sum_{k}c_{k}$):

\begin{equation}
\Psi^{R}(\mathbf{r},t)=\widetilde{\Psi}^{R}(\mathbf{r},t',t'',t)|_{t''=t'=t}=e^{-\frac{i}{\hbar}\int^{t}d\tau
E^{ad}_{j}(\tau)}\phi^{ad,R}_{j}(\mathbf{r},t',t'')|_{t''=t'=t}\label{eq54}
\end{equation}
This is the adiabatic solution of the TDSE associated with initial
state $\varphi_{j}(\textbf{r})$.

Let us summarize and clarify the procedure that need to be made in
order to determine if a pulse is adiabatic or not; The
determination is carried out through the calculation of the
expression $A_{(j,0)}(t)$ (it is t now, not t'' !):

(1) Perform non-hermitian adiabatic Floquet simulations
(Eq.\ref{eq27}) with \textbf{CW} field, for a range of intensities
$\varepsilon_{1}$ which covers all intensities between zero and
the maximal intensity $\varepsilon_{0}$ that the studied laser
pulse reaches. The adiabatic Floquet Hamiltonian is therefore

\begin{equation}
H_{ad}(\mathbf{r},t',\varepsilon_{1}) \equiv  H_{0}(\mathbf{r})-e
\mathbf{r} \cdot \varepsilon_{1} cos(\omega t')-i\hbar
\frac{\partial}{\partial t'}
 \label{eq48}
\end{equation}
and the eigenvalue equation to be solved is

\begin{equation}
H_{ad}(\mathbf{r},t',\varepsilon_{1})
\phi^{ad,R}_{j}(\mathbf{r},t',\varepsilon_{1})=E^{ad}_{j}(\varepsilon_{1})
\phi^{ad,R}_{j}(\mathbf{r},t',\varepsilon_{1})
 \label{eq49}
\end{equation}
Obtain the quasienergy spectrum $E^{ad}_{j}(\varepsilon_{1})$ and
all the adiabatic cross terms
$c_{(j',m)}^{(j,0)}(\varepsilon_{1})$ as defined in
Eq.\ref{eq45},\ref{eq49} as function of the intensities. This
stage is done only once, for a given system.

(2) For a given laser pulse $f(t)$ with a maximal intensity
$\varepsilon_{0}$, evaluate the effective CW-field intensity as
function of time $\varepsilon_{0}f(t)$ (Eq.\ref{eq18}). Then,
convert the adiabatic cross terms to be functions of time via the
equality (Eq.\ref{eq26a})

$$
\varepsilon_{1}=\varepsilon_{0}f(t).
$$
using $c_{(j',m)}^{(j,0)}(\varepsilon_{1})$ as calculated in step
1
$c_{(j',m)}^{(j,0)}(t)=c_{(j',m)}^{(j,0)}(f^{-1}(\varepsilon_{1}/\varepsilon_{0}))$
where $f^{-1}$ is the transformation which fulfills
$f^{-1}[f(t)]=t$.

(3) For each given resonance state $\alpha=(j,0)$ calculate

\begin{equation}
A_{(j,0)}(t)=F(t)\cdot \sum_{j'\neq j} \sum_{m\neq 0}
|c_{(j',m)}^{(j,0)}(t)|
 \label{eq49a}
\end{equation}
using $c_{(j',m)}^{(j,0)}(t)$ from step 2 where notice that here
$t''$ in Eqs.(\ref{eq42}-\ref{eq45}) is replaced by $t$. If for a
given resonance state $\alpha=(j,0)$ the corresponding expression
$A_{(j,0)}(t)$ is smaller then unity at every instant, it is
guaranteed that the system initially at the bound state which
corresponds to this resonance, will evolve adiabatically to that
resonance. In this case the HGS spectra will show only odd
harmonics and the ionization probability as function of time will
have a simple form that will be shown.

\section{Illustrative numerical example}
We studied a single-electron 1D Xe atom subjected to a single
sin-square pulse of strong monochromatic laser field in two
approaches. In the first one hermitian simulations were carried
out whereas in the second non-hermitian Floquet simulations based
on the complex scaling method were carried out. The hermitian
simulations were carried out by solving the following TDSE

\begin{equation}
[-\frac{\hbar^{2}}{2m} \frac{\partial^{2}}{\partial
x^{2}}+V_{0}(x)-ex \varepsilon_{0}f(t)cos(\omega
t)]\Psi(x,t)=i\hbar \frac{\partial}{\partial t}\Psi(x,t)
\label{eq55}
\end{equation}
with a sine-square envelope

\begin{equation}
f(t)=\sin(\frac{\omega t}{2 N}) \label{eq56}
\end{equation}
and the field-free effective potential $V_{0}(x)$ was an inverse
Gaussian

\begin{equation}
V_{0}(x)=-0.63\exp(-0.1424x^{2}) \label{eq57}
\end{equation}
which supports two bound states that mimic the two lowest
electronic states of Xe, with energies $E_{0}=-0.4451
\textit{a.u.}$, $E_{1}=-0.1400 \textit{a.u.}$ and a third weakly
bound state with energy $E_{2}=-0.00014 \textit{a.u.}$

The wave function was taken initially at the ground state (g.s.)
of the field-free Hamiltonian of the system:

\begin{equation}
\Psi(x,t=0)=\varphi_{1}(x) \label{eq58}
\end{equation}
and was calculated for times $0<t<NT$ (single sine-square pulse,
$N$ was the number of optical cycles that entered the pulse).
Using this wavefunction the ionization probability at times
$t=\frac{\pi}{2 \omega}+n\frac{\pi}{\omega}, 0\leq n\leq 2N-1$ was
calculated (at these times the potential felt by the electron was
the field-free potential $V_{0}(x)$)

\begin{equation}
P_{ion}(t)=1-\sum_{i=1}^{3}|\langle\varphi_{i}|\Psi(t)\rangle|^{2}
\label{eq59}
\end{equation}
where $\varphi_{i}$ were the bound states of the field-free
problem (3 bound states over which we summed in this example).
Also HGS was calculated, which following the classical-quantum
correspondence principle (Larmor formula,\cite{L&L}) equals the
modulus-square of the Fourier-transformed time-dependent
acceleration expectation value. This is actually the intensity of
the radiation emitted by the oscillating electron as presented in
energy space.

\begin{equation}
\sigma(\Omega)= |\frac{1}{NT}\int_{0}^{NT} \frac{1}{m}\langle
\Psi(t)|-\frac{\partial V_{0}(x)}{\partial x}+e \varepsilon_{0}
f(t) cos(\omega t)|\Psi(t)\rangle~~ e^{-i\Omega t } dt |^{2}
\label{eq60}
\end{equation}

The non-hermitian simulations were Floquet simulations which were
carried out for different field intensities. The quasienergy
spectrum of complex energies $E^{ad}_{j'}(\varepsilon_{1})$ and
the adiabatic cross terms $c_{(j',m)}^{(j,0)}(\varepsilon_{1})$
were calculated for each intensity. Then, these quantities were
expressed as function of time through Eq.\ref{eq26a}, i.e.
$t=f^{-1}(\varepsilon_{1}/\varepsilon_{0})$. It was verified that
in the cases where the adiabatic criteria was fulfilled, the HGS
obtained from the hermitian propagation simulation contained only
odd harmonics. A more quantitative measure of the existence of the
adiabatic criteria was obtained through the comparison of
ionization probabilities as obtained from the hermitian
propagation simulation and the non hermitian simulation.

In order to get non-hermitian Floquet Hamiltonian, the complex
coordinate method was used. The Floquet Hamiltonian
(Eq.\ref{eq48})

\begin{equation}
H_{ad}(x,t',\varepsilon_{1}) \equiv
-e^{-2i\theta}\frac{\hbar^{2}}{2m} \frac{\partial^{2}}{\partial
x^{2}}+V_{0}(x e^{i\theta})-ex e^{i\theta}
\varepsilon_{0}f(t)cos(\omega t) -i\hbar \frac{\partial}{\partial
t'} \label{eq60a}
\end{equation}
was diagonalized. Provided that the scaling parameter $\theta$ was
sufficiently large, the resonance quasienergy states were $\theta$
independent:

\begin{equation}
E_{j}^{ad}(t'')=E_{r_{j}}(t'')-\frac{i}{2}\Gamma_{j}(t'')~~;~~\Gamma_{j}=\frac{\hbar}{\tau_{j}}\label{eq62}
\end{equation}
$E_{r_{j}}$ being the position of the state, $\Gamma_{j}$ being
the width of the state and $\tau_{j}$ its lifetime. Since the
resonance states had finite lifetimes, and since the resonances
are the states which are associated with the dynamics, these
resonance lifetimes should have fingerprints in the hermitian
propagation simulation. Indeed it was found to be so when the
ionization probabilities, as computed in the two simulations, were
compared; The ionization probability at each instant was given by
the following expression, which was obtained using Eq.\ref{eq54}
and the F-product definition for the inner product
\cite{F-product}:

\begin{equation}
P_{ion}(t)=e^{-\frac{1}{\hbar}\int^{t}d\tau \Gamma_{1}(\tau)}
\label{eq63}
\end{equation}
where $\Gamma_{1}(\tau)$ was associated with the Floquet
resonanace state that was "born" from the ground stationary state
$\varphi_{1}(\textbf{r})$ as the field was turned on. The
resonances that were "born" from the field-free Hamiltonian bound
states were identified by plotting the quasienergy spectrum as
function of the effective field intensity $\varepsilon_{1}$. The
resonance complex quasienergy trajectories started from the
field-free Hamiltonian bound states real energies and formed
continuous trajectories in the complex energy plane as function of
the intensity.

How was this expression for the ionization probability obtained?
According to the F-product definition when the complex energy
given in Eq.\ref{eq62} is substituted in Eq.\ref{eq54}, the "ket"
(right) solution of the non-hermitian TDSE is obtained:

\begin{equation}
\Psi^{R}(\mathbf{r},t)=e^{-\frac{1}{2 \hbar}\int^{t}d\tau
\Gamma_{j}(\tau)~-\frac{i}{\hbar}\int^{t}d\tau
E_{r_{j}}(\tau)}\phi^{ad,R}_{j}(\mathbf{r},t',t'')|_{t''=t'=t}\label{eq64}
\end{equation}
and it is easily seen that this function decays with time. The
"bra" (left) solution is not a solution of a Schr\"{o}dinger
equation but is derived from the "ket" solution (the explanation
of this point is beyond the scope of this work; for an explanation
see \cite{F2-product})

\begin{equation}
\Psi^{L}(\mathbf{r},t)=e^{-\frac{1}{2 \hbar}\int^{t}d\tau
\Gamma_{j}(\tau)~+\frac{i}{\hbar}\int^{t}d\tau
E_{r_{j}}(\tau)}\phi^{ad,L}_{j}(\mathbf{r},t',t'')|_{t''=t'=t}\label{eq65}
\end{equation}
and also this function decays with time. When the F-product of
$\Psi^{L}(\mathbf{r},t)$ and $\Psi^{R}(\mathbf{r},t)$ is
calculated (this is an overlap integral without complex
conjugation of the left state; in the usual dirac-product notation
this reads $\langle \Psi^{L*}(t)|\Psi^{R}(t) \rangle$) the terms
containing the real part of the energy cancel each other and the
overlap integral (c-product) of the adiabatic Floquet states gives
unity (remember the completeness property of Floquet states also
in coordinate space alone). The expression in Eq.\ref{eq63} is
obtained for the specific case that only the Floquet resonance
state which is "born" from the field-free bound state is
populated. When several Floquet states are populated the same type
of calculation could be repeated, where this time
$\Psi^{L}(\mathbf{r},t)$ and $\Psi^{R}(\mathbf{r},t)$ are given by
liner combination of terms as in Eq.\ref{eq64} with the proper
coefficients. Since the adiabatic Floquet states are orthogonal to
each other, a generalization of the result of Eq.\ref{eq63} is
obtained:

\begin{equation}
P_{ion}(t)=\sum_{i=1}^{3}|\langle\varphi_{i}|\Psi(t=0)\rangle|^{2}e^{-\frac{1}{\hbar}\int^{t}d\tau
\Gamma_{i}(\tau)} \label{eq65a}
\end{equation}
Notice that in the non-hermitian formalism the norm is not
conserved but decays with time; it contains the knowledge about
decay processes inherently.

The two functions as given in Eq.\ref{eq59},\ref{eq63} were
compared and it was found that the resemblance between the
functions increased as the adiabatic limit was increasingly
reached by the laser pulse parameters.

The numerical method used to solve the TDSE with hermitian
Hamiltonian was the split operator Forest-Ruth algorithm with 7
points \cite{Split}; The grid size, time step and/or grid step
were adjusted as required to achieve convergence. For the
non-hermitian Floquet simulation, the (t,t') formalism was used,
with the complex coordinate method. The number of basis functions,
box length and scaling angle were adjusted as required to achieve
convergence.

\begin{figure}[h]
\centerline{ \epsfxsize=14cm \epsfysize=14cm
\epsffile{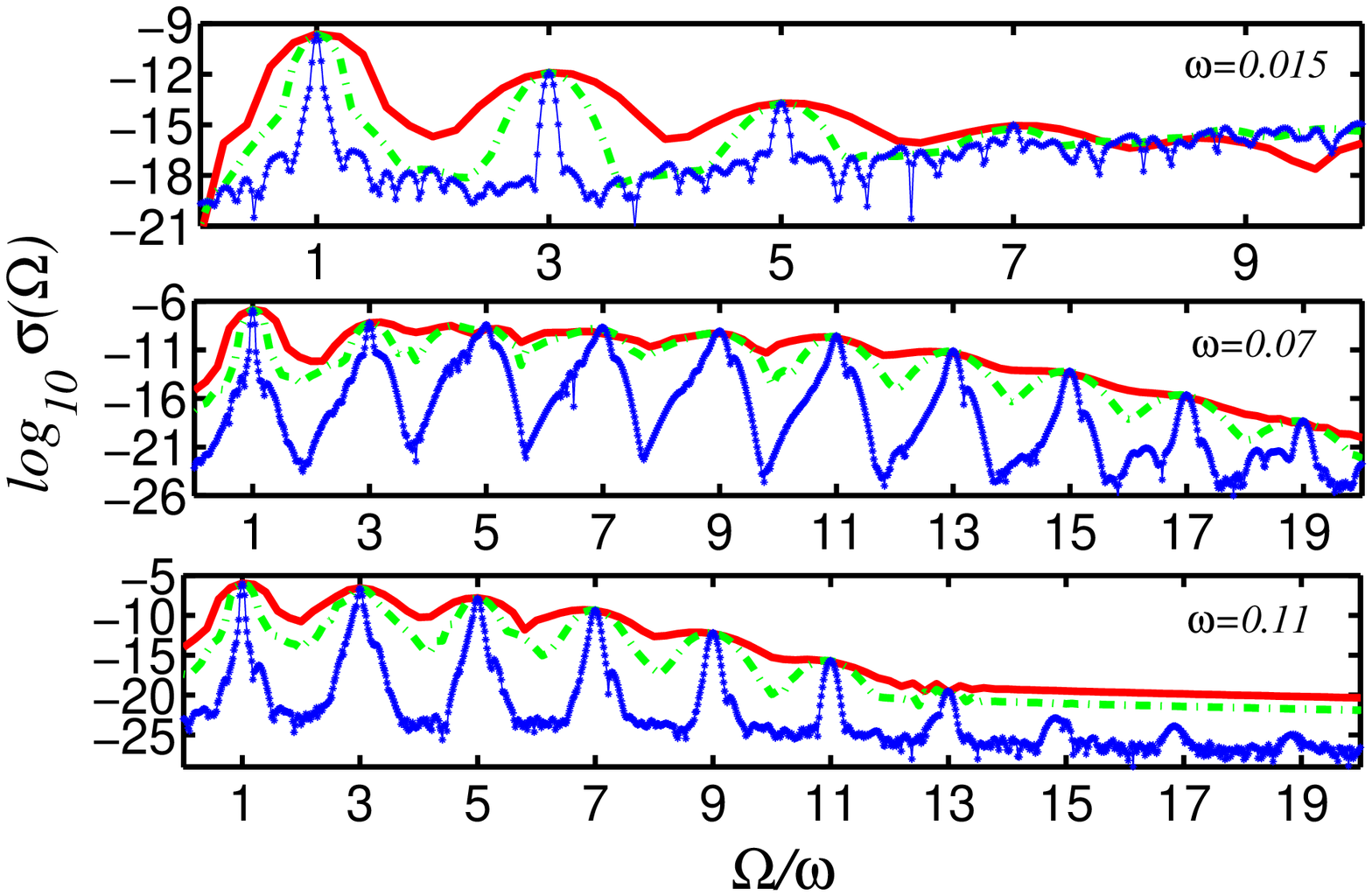}} \caption{\label{fig1} {(color
online) HGS obtained from hermitian simulations describing a 1D Xe
atom subjected to various sin-square pulses with strength of
$\varepsilon_{0}=0.035a.u.$, laser frequencies of
$\omega=0.015a.u.$ (upper part), $\omega=0.07a.u.$ (middle part)
and $\omega=0.11a.u.$ (lower part) and pulse-durations of $N=5$
(solid), $10$ (dot-dashed) and $50$ (solid-dotted) optical cycles.
The initial state was taken to be the ground state in all
simulations. For the longest pulse duration only odd harmonics
appear in the spectrum, and hyper Raman lines are absent. This is
also the basic feature in the case of short pulses, although some
deviation from this structure appear at frequency of
$\omega=0.07a.u.$. }}
\end{figure}

In Fig.\ref{fig1} the HGS as obtained for pulse strength of
$\varepsilon_{0}=0.035a.u.$ (corresponding to intensity of
$4.30\cdot 10^{13}W/cm^{2}$), pulse-durations of $N=5, 10, 50$
optical cycles, laser frequencies of $\omega=0.015a.u., 0.07a.u.,
0.11a.u.$ (corresponding to energies of $\hbar\omega=0.408eV,
1.90eV, 2.99eV$ and wavelengths of $\lambda=3040nm, 651nm$,
$414nm$, respectively) and initial state $\varphi_{1}(x)$ are
shown. The appearance of odd harmonics and the absence of even or
non-integer harmonics, is in general the main feature which
appears for long pulses, regardless of the frequency. The odd
harmonics are obtained even for the shortest laser pulses although
the existence of odd-symmetry selection rules is sensitive to the
frequency: some deviation appears at frequency of
$\omega=0.07a.u.$ and the obtained spectra is more complicated.

\begin{figure}[h]
\centerline{ \epsfxsize=14cm \epsfysize=14cm
\epsffile{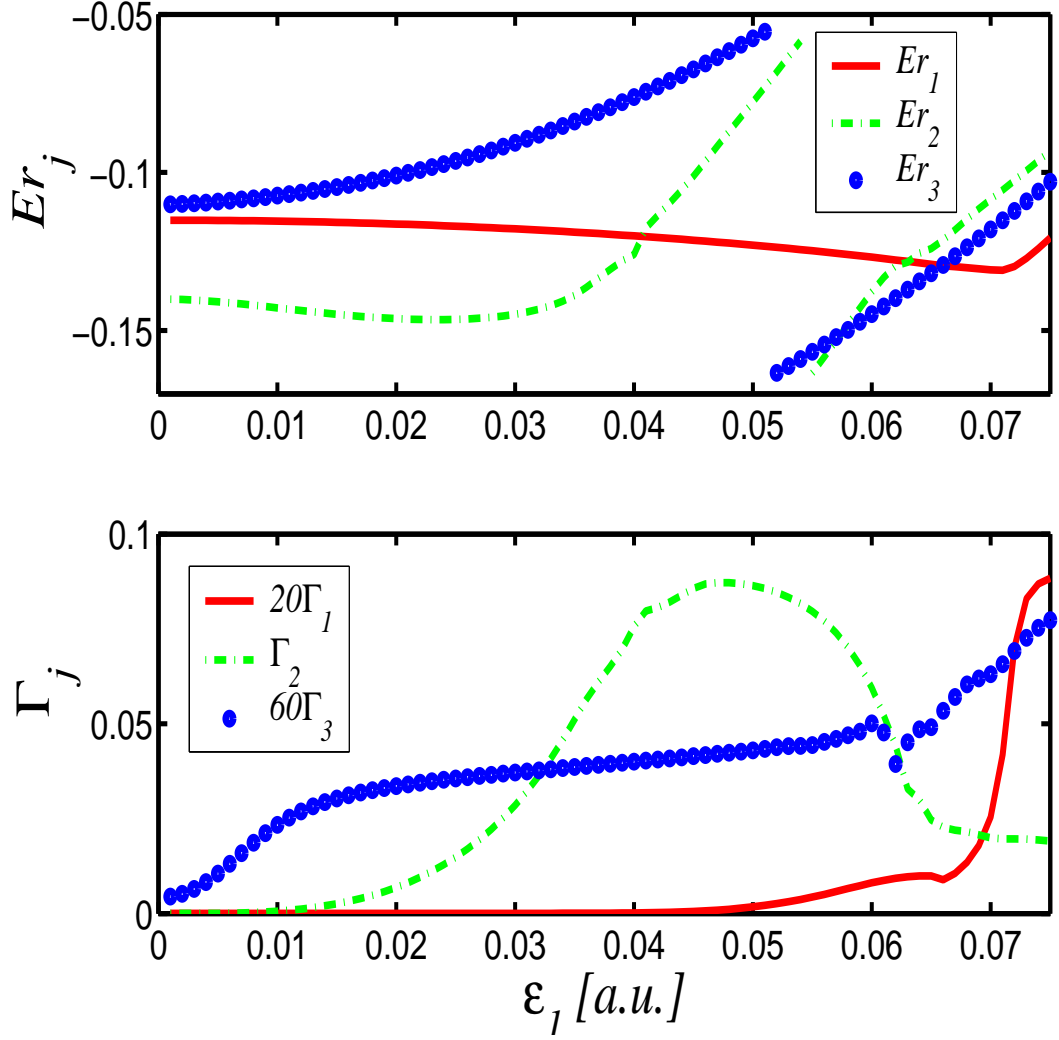}} \caption{\label{fig2} {(color
online) Positions (upper part) and lifetimes (lower part) of the
complex quasienergies of resonances 1-3 as function of the
CW-field strength $\varepsilon_{1}$ as obtained from adiabatic
non-hermitian Floquet simulations (Eq.\ref{eq49},\ref{eq62}) for
$\omega=0.11a.u.$. }}
\end{figure}

In Fig.\ref{fig2} the complex quasienergies of the 3 resonances
which are "born" from the 3 bound states of the field-free
Hamiltonian (will be given indices 1-3 from now on), as obtained
from adiabatic Floquet simulations (Eq.\ref{eq49}) for
$\omega=0.11a.u.$ are shown as function of the CW-field strength
$\varepsilon_{1}$. Notice that the lifetimes of these resonances
are not monotonic functions of the field intensity. It should be
noted that these were not the only resonances that appeared in the
quasienergy spectrum; there were also other resonances which
didn't emerge from the field-free bound states. However, as will
be shown in Fig.\ref{fig4}, in this case for not too large field
intensities resonances 1-3 were the only important resonances and
they alone determined the dynamical behavior of the system.

\begin{figure}[h]
\centerline{ \epsfxsize=14cm \epsfysize=14cm
\epsffile{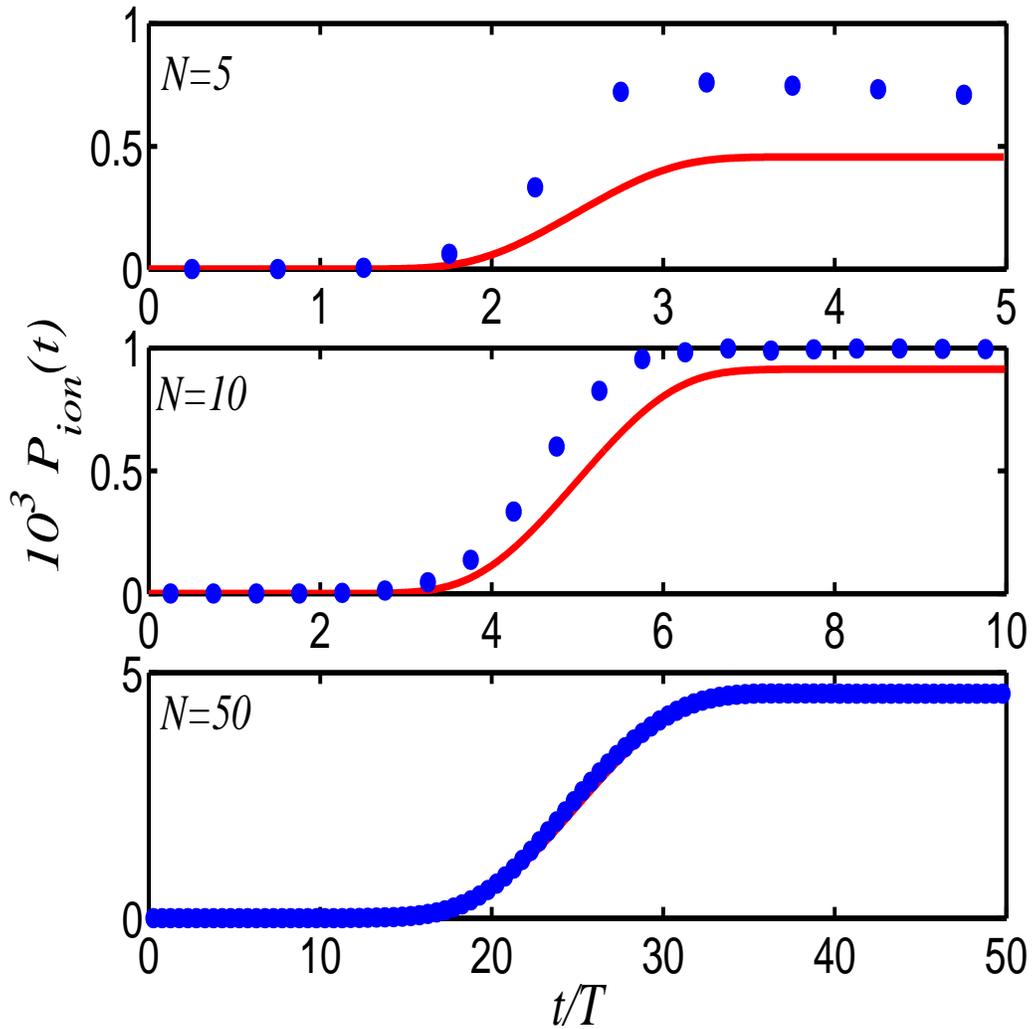}} \caption{\label{fig3} {(color
online) Ionization probabilities obtained from hermitian
simulations (lines) (Eq.\ref{eq59}) for pulse-durations of $N=5$
(upper part), $N=10$ (middle part), $N=50$ (lower part) optical
cycles, and from non-hermitian simulations (dots) (Eq.\ref{eq63}).
As the pulse becomes longer the results obtained from the two
simulations become identical since the adiabaticity of the process
is increased. }}
\end{figure}

\begin{figure}[h]
\centerline{ \epsfxsize=14cm \epsfysize=14cm
\epsffile{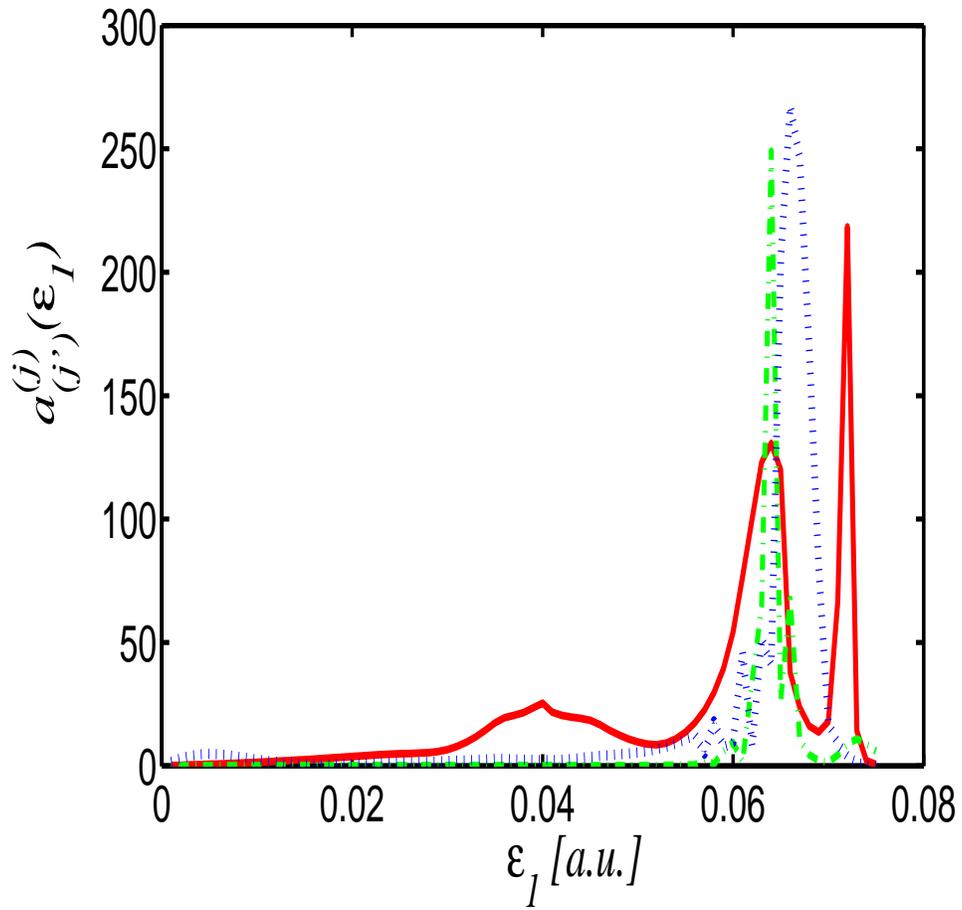}} \caption{\label{fig4} {(color
online) The expressions $6a_{(2)}^{(1)}(\varepsilon_{1})$ (solid),
$a_{(3)}^{(1)}(\varepsilon_{1})$ (dashed) ,
$2.5a_{(3)}^{(2)}(\varepsilon_{1})$ (dotted) (Eq.\ref{eq44}) as
function of the CW-field strength $\varepsilon_{1}$, for
$\omega=0.11a.u.$. }}
\end{figure}

In Fig.\ref{fig3} the ionization probability as obtained from the
hermitian simulation (Eq.\ref{eq59}) with pulse strength of
$\varepsilon_{0}=0.035a.u.$, pulse-durations of $N=5, 10, 50$
optical cycles, laser frequency of $\omega=0.11a.u.$ and initial
state $\varphi_{1}(x)$, is compared to the ionization probability
as obtained from the expression given in Eq.\ref{eq63}, which is
derived from the F-product formalism together with the resonances
quasienergies obtained from the non-hermitian simulation. It is
seen that as the pulse becomes longer the results obtained from
the two simulations become identical.

Both results of HGS and ionization probabilities showed that the
time-dependent wavefunction of the studied systems could be well
approximated by the adiabatic expression given in Eq.\ref{eq54},
even for short pulses. The values of the terms which check the
adiabatic criteria, as seen in Fig.\ref{fig6}, gave the
explanation why this was so.

In Fig.\ref{fig4} the expressions $a_{(2)}^{(1)}, a_{(3)}^{(1)},
a_{(3)}^{(2)}$ (Eq.\ref{eq44}) which describe the couplings
between every two resonances from the set of 3 tracked resonances
is shown as function of the CW-field strength $\varepsilon_{1}$,
for the case $\omega=0.11a.u.$. It can be seen that the coupling
between resonances 1 and 2 is strong for field strength of
$\varepsilon_{1}\simeq 0.04a.u.$, $0.064a.u.$ and $0.072a.u.$.
This could be partially explained on the basis of the values of
the quasienergies, as seen in Fig.\ref{fig2}, at least for two
field strengths out of the three. It is seen that for field
strength of $0.04a.u.$ and $0.064a.u.$ the real parts of the
quasienergies cross, resulting in small value of the denominator
in the expression given in Eq.\ref{eq45} for $c_{(2,m)}^{(1,0)}$
(at least for one $m$ term). In the same way the strong couplings
between resonances 1 and 3 at field strength of
$\varepsilon_{1}=0.064a.u.$, and between resonances 2 and 3 at
field strength of $\varepsilon_{1}=0.066a.u.$ could be explained
on the basis of the quasienergy values at these field-strengths.
In particular it should be noticed that the couplings between
resonances 1 and 3 at $\varepsilon_{1}=0.064a.u.$ are the
strongest among all 3 resonances due to the close values of both
real and imaginary parts of the quasienergies. However, it should
be noted that crossings in the quaisenergy plot are not always
indications of large couplings since also the overlap between the
wavefunctions (the nominator of the expression for
$c_{(j',m)}^{(j,0)}$) is important.

\begin{figure}[h]
\centerline{ \epsfxsize=14cm \epsfysize=14cm
\epsffile{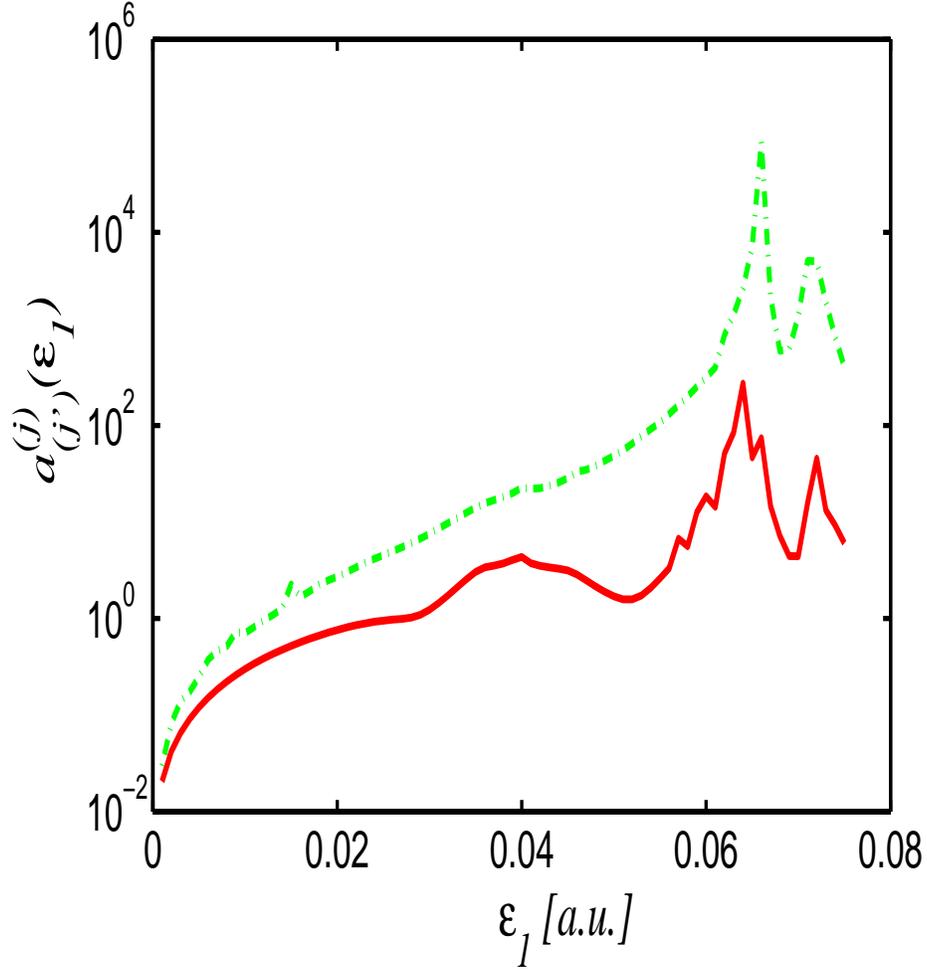}} \caption{\label{fig5} {(color
online) The expressions $\sum_{j'\neq
1}a_{(j')}^{(1)}(\varepsilon_{1})$ (dashed) and
$a_{(2)}^{(1)}(\varepsilon_{1})+a_{(3)}^{(1)}(\varepsilon_{1})$
(solid), which describe respectively the couplings between
\textbf{all} quasistates of the system $\alpha'=(j',m)$ or only
resonances 2 and 3 to resonance 1 $\alpha=(1,0)$ are shown as
function of the CW-field strength $\varepsilon_{1}$, for the case
$\omega=0.11a.u.$. }}
\end{figure}

\begin{figure}[h]
\centerline{ \epsfxsize=14cm \epsfysize=14cm
\epsffile{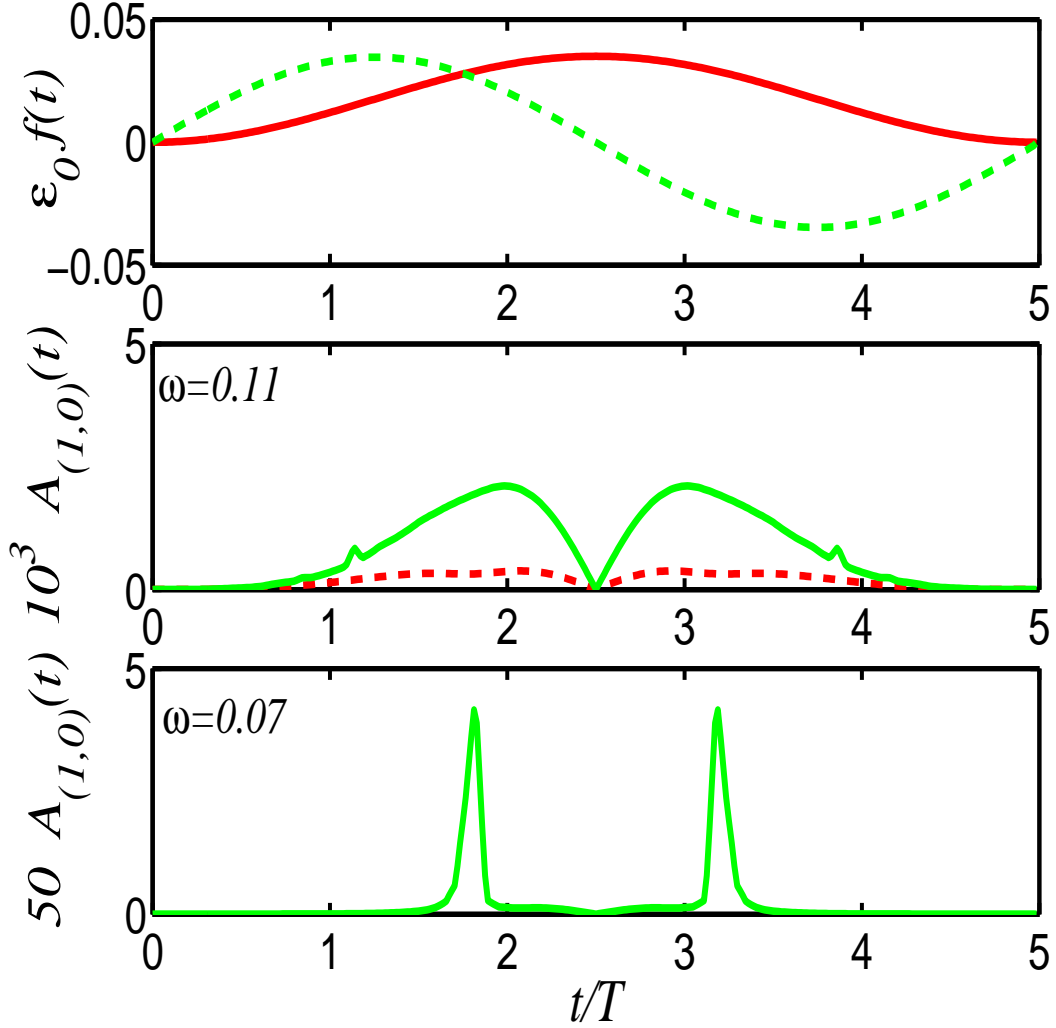}} \caption{\label{fig6} {(color
online) upper part: The time-dependence of a sine-square pulse
$\varepsilon_{0} sin^{2}(\frac{\omega t}{2N})$ (solid) and a
scaled time derivative
$90(\frac{\varepsilon_{0}\omega}{2N})sin(\frac{\omega t}{N})$
(dashed) for $N=5$, $\omega=0.11a.u.$ and
$\varepsilon_{0}=0.035a.u.$. Middle part: the full term
$A_{(1,0)}(t)$ (solid), and the partial term
$F(t)(a^{(1)}_{(2)}(t)+a^{(1)}_{(3)}(t)$ (dashed) are shown as
function of time. Both terms are bounded by the small value of
$\sim 2\cdot 10^{-3}$ for all times. Lower part: the same as in
the middle part, but for $\omega=0.07a.u.$. Both terms are bounded
by a larger value of $\sim 10^{-1}$. It is therefore deduced that
for $\omega=0.11a.u.$ the process is adiabatic but for
$\omega=0.07a.u.$ it is not and this is indeed verified in the HGS
given in Fig.\ref{fig1} for $N=5$. }}
\end{figure}

In Fig.\ref{fig5} the sum $\sum_{j'\neq
1}a_{(j')}^{(1)}(\varepsilon_{1})$ which describes the couplings
between \textbf{all} quasistates of the system $\alpha'=(j',m)$ to
resonance 1 $\alpha=(1,0)$ is shown as function of the CW-field
strength $\varepsilon_{1}$, for the case $\omega=0.11a.u.$. In
addition, also the partial sum
$a_{(2)}^{(1)}(\varepsilon_{1})+a_{(3)}^{(1)}(\varepsilon_{1})$
which describes the couplings between resonances 2 and 3 to the
resonance state $\alpha=(1,0)$ is shown. It can be shown that up
to a moderate intensity of $\varepsilon_{1}\sim 0.04a.u.$ the
first resonance is mainly coupled only to the other 2 resonances
and not to other, higher resonances or continuum states. The
entire dynamics is governed almost solely by the 3 resonances
which are "born" from the 3 bound states of the field-free
Hamiltonian.

The sums described above, which are functions of a CW-field
strength $\varepsilon_{1}$, are converted to be explicit functions
of time for the specific sine-square pulse with maximal intensity
$\varepsilon_{0}=0.035a.u.$ used in the simulation
(Fig.\ref{fig5}, upper part). For this purpose for each time $t$
$0<t<NT$ the effective CW-field strength $0.035f(t)$ is calculated
and the values of the 2 functions shown in Fig.\ref{fig5} which
fit this effective CW-field strength are taken. Hence the 2 sum
functions are converted to be explicit functions of time. In the
middle part of Fig.\ref{fig6} the full term $A_{(1,0)}(t)$, which
represents the degree of adiabaticity in the process of shining a
1D Xe atom initially at the ground state with a sine-square laser
pulse supporting $N=5$ optical cycles of monochromatic radiation
with frequency $\omega=0.11a.u.$ and strength
$\varepsilon_{0}=0.035a.u.$, is shown as function of time. It is
seen that the term $A_{(1,0)}(t)$ is bounded by the value of $\sim
2\cdot 10^{-3}$ for all times, whether it is calculated by
coupling of the first resonance to other 2 resonances only or to
all other states. The structure of the HHS for $N=5$ seen in
Fig.\ref{fig1} implies that this value is indeed small and the
process is adiabatic.  In the lower part of Fig.\ref{fig6} the
same quantity is shown, but for $\omega=0.07a.u.$. Here it is seen
the terms $A_{(1,0)}(t)$ are bounded by a much larger value of
$\sim 10^{-1}$ for all times and the more complex structure of the
HHS for $N=5$ seen in Fig.\ref{fig1} implies that this value is
not small enough to indicate the appearance of an adiabatic
process. It should be noted that for a given system with
field-strength $\varepsilon_{0}$, frequency $\omega$ and
sine-square pulse envelope, we have $F(t)=|e|(\frac{\hbar
\varepsilon_{0}\omega}{2N})|sin(\frac{\omega t}{N})|$. Therefore,
as the number of optical cycles the pulse supports increases the
general shape of the terms $A_{(1,0)}(t)$ is kept the same but is
attenuated. Therefore, for $\omega=0.07a.u.$ as $N$ gets bigger,
the pulse's envelope varies more slowly and the process becomes
more and more adiabatic, as seen in the HHS spectra for $N=50$ for
example.

\section{Conclusions}

With the help of the (t,t') formalism we derive an adiabatic
theorem for open systems. The use of the complex scaling
transformation plays a key role in our derivation. For example,
the spectrum of the Floquet Hamiltonian of an open system is
changed dramatically. Rather than a continuous spectrum that is
responsible for the absence of an adiabatic limit for $N$(number
of basis functions)$\rightarrow \infty$ in the conventional QM,
the resonances are associated with a point spectrum and are
separated from the continuum which is rotated into the lower half
of the complex energy plane.

An interesting important numerical result of our derivation is
that the calculation of the effect of the pulses's shape on the
dynamics does not require heavy computations. The entire effect of
the laser pulse is embedded in a multiplication factor of
$df(t)/dt$ where $\varepsilon_0 f(t)$ is the variation of the
maximum field amplitude as function of time.

As a numerical example we applied  the adiabatic theorem we
derived to a model Hamiltonian of Xe atom (with symmetric
field-free potential) which interacts with strong, monochromatic
laser pulses. We have shown that the generation of odd-order
harmonics and the absence of even-order harmonics, even when the
pulses are extremely short, can be explained with the help of the
adiabatic theorem we derived.

The use of a single-electron 1D model to describe a realistic atom
is justified since it has been shown before in many cases that all
the main strong field effects are reproduced. Therefore the
conclusions obtained with this model are also valid for a
realistic atom.

\section{Acknowledgements}
 This work was supported in part by the by
the Israel Science Foundation
 and by the Fund of promotion of research at the Technion. Dr. Milan \v{S}indelka
is acknowledged for  most helpful and fruitful discussions.

\newpage
\bibliographystyle{plain}

\newpage


\end{document}